\documentclass{article}
\usepackage{epsfig}
\usepackage{graphics}
\usepackage{float}
\usepackage{amsmath}

\newcommand{\be}{\begin{equation}}
\newcommand{\ee}{\end{equation}}

\begin{document}

\title{Screening magnetic fields by a superconducting disk: a simple model}

\author{ J.-G. Caputo$^{1}$,
L. Gozzelino$^{2}$, F.Laviano$^{2}$, G.Ghigo$^{2}$, R.Gerbaldo$^{2}$, \\
J.Noudem$^{3}$, Y.Thimont$^{3}$  and P.Bernstein$^{3}$
}

\maketitle

{\normalsize \noindent
$^1$ : Laboratoire de Mathematiques,\\
INSA de Rouen,\\
Avenue de l'Universite,\\
76801 Saint-Etienne du Rouvray, France \\
E-mail: caputo@insa-rouen.fr \\
$^2$ : Department of Applied Science and Technology, \\
Politecnico di Torino\\
Torino, Italy.\\
$^3$: CRISMAT , \\Physics Department, \\
Universit\'e de Caen, France
}

\date{\ }

\begin{abstract}
We introduce a simple approach to evaluate the magnetic field
distribution around superconducting samples, based on the London
equations; the elementary variable is the vector potential. This
procedure has no adjustable parameters, only the sample geometry
and the London length, $\lambda$, determine the solution. The
calculated field reproduces quantitatively the measured induction field
above MgB$_2$ disks of different diameters, at 20K and for applied
fields lower than 0.4T. The model can be applied if
the flux line penetration inside the sample can be neglected
when calculating the induction field distribution
outside the superconductor. Finally we show on a cup-shape
geometry how one can design a magnetic
shield satisfying a specific constraint.
\end{abstract}

\section{Introduction}

Magnetic field screening is very important for a large variety
of applications. Very low magnetic field background are required
when high resolution magnetic field detector are used
(e.g. SQUID \cite{Xia, Korber}). Magnetic shielding is also required
to solve problems of electromagnetic compatibility among different
devices (i.e., the simultaneous use of multiple diagnostic
devices including the magnetic resonance imaging \cite{Delso})
or for military applications \cite{Baltag}.

Depending on the application, active \cite{Harrison} or passive
\cite{Seki, Denis192} shielding solutions can be adopted. In
static or quasi-static regimes passive shielding can be
achieved using ferromagnetic and/or superconducting materials.
The former, but not the latter, can operate at room temperature.
However the latter, due to the Meissner effect, show the
highest shielding efficiency.

For type-II superconductors, complete magnetic shielding
occurs only when the
total field is below the value of the lower critical field,
$B_{c1}$. Here we disregard the region of depth $\lambda$ - the 
London  penetration depth - where shielding currents are confined.
If the applied field is much larger than $B_{c1}$,
a description of the magnetic field of the
superconductor cannot disregard the vortex penetration and
movement inside the materials. Several experiments of magnetic shielding
have been carried out in the last years, using both low-Tc and
high-Tc superconducting materials operating in the mixed state.
In this state, the interpretation of the experimental results requires
to calculate the flux lines distribution inside and outside the sample.
One needs models such as the critical state model
\cite{bean, brandt98, sanchez01} associated with a constitutive law
giving the non-linear dependence of the electric field on the
current density to account for the dissipation due to vortex
motion \cite{Denis192, denis418, fagnard}. Because of this
complexity, this approach yields exact solution only in
few idealized cases \cite{Navau13}.

In addition to the material, another important issue
to meet the different application requirements, is to
produce magnetic shields with more complex geometries.
Moreover, an approach to calculate easily
how the shield geometry influences the field
distribution outside the sample is the first step
towards solving the inverse problem of designing
a magnetic shield starting from given requirements.

Aiming at this, we introduce an approach
based on the London equations, where the elementary variable
is the vector potential $\mathbf{A}$ \cite{tinkham,vanduzer}. The
order parameter is assumed constant throughout the sample leading
to a very simple London equation for $\mathbf{A}$. There the
medium is represented by a source term. This formulation
guarantees the continuity of the vector potential and gives
in a simple way the magnetic induction field everywhere. In particular,
it allows us to study in detail the field outside the sample
and to take into account easily the demagnetization field. This
model is strictly valid only for applied fields below
B$_{c1}$. However, it is possible to extend its application
range also for magnetic fields larger than the lower
critical one, provided that the field penetration inside the sample
can be neglected when evaluating the magnetic field distribution
outside the sample itself. This is verified as long as
the magnetic field amplitude outside
the sample scales linearly with the applied field
as expected from the theory.

This approach was validated by comparing the calculation outputs
with the experimental results obtained on three MgB$_2$ disks
with different aspect ratio. The numerical results are in
quantitative agreement with the measures for applied fields lower
than 0.4T. The choice of this superconducting material
was made on the basis of its numerous advantages. First of all,
its working temperature (10-30 K \cite{Tomsic}) can be
easily reached using one-stage cryogen free cryocoolers. Then
this material shows higher B$_{c1}$ and coherence length, $\xi$,
than high-Tc cuprates. This last property ensures the transparency
of grain boundaries to current flow \cite{glb02} and, as
a consequence, the possibility both to work with polycrystalline
samples and to produce specimens with complex shapes
assembled by soldering elementary pieces
\cite{Mikheenko, soudure}. Finally, the low density value
of MgB$_2$ makes this material a good candidate for applications
where weight constraints are present. 

The article is organized as follows. In section II, we derive
the model from first principles and show how it is solved. Section III
describes the fabrication of the samples and the experimental details of the characterization. The experimental data are
presented and discussed in comparison with the model
predictions in section IV. Section V shows how a practical
magnetic screen can be designed based on a quantitative criterion.

\section{The model}

The Maxwell equations of magnetostatics are
\be\label{maxwell}
\nabla\cdot\mathbf{B}=0,~~\nabla\times \mathbf{B} = \mu_0 \mathbf{J},
\ee
where $\mathbf{B}$ is the magnetic induction field and where $\mathbf{J}$ is the current density.
The electric field is omitted because we are considering the superconductor
only in the Meissner state.
We introduce the vector potential $\mathbf{A}$ such that
$$\mathbf{B} = \nabla\times\mathbf{A}.$$
Taking the curl of the second equation in (\ref{maxwell}) we get
\be\label{maxaj}-\nabla^2 \mathbf{A} = \mu_0 \mathbf{J},\ee
where we have assumed the London gauge
$$\nabla\cdot \mathbf{A} =0.$$
The London hypothesis, i.e. there is no phase momentum in the
superconductor \cite{vanduzer} implies
\be\label{london} \mu_0 \mathbf{J} = - {1 \over \lambda^2} \mathbf{A},\ee
where $\lambda$ is the London penetration depth.
Combining equations (\ref{maxaj}) and (\ref{london}) we get
\be\label{lmaxa1}
\nabla^2 \mathbf{A} = {1 \over \lambda^2} \mathbf{A}.\ee
Note that the current $\mathbf{J}$ only exists in the superconductor, outside
it is zero. The equation can then be written so it describes the
field everywhere inside and around the superconductor. It reads
\be\label{lmaxa}
\nabla^2 \mathbf{A} = {1 \over \lambda^2} \mathbf{A} I({\bf r}),\ee
where $I({\bf r})=0$ (resp. $I({\bf r})=1$) outside (resp. inside)
the superconductor.

This equation is a first order description of the superconductor in
the sense that we assumed the order parameter $\Phi$ to be
spatially uniform, i.e. the superconductor is in the Meissner
state. To see this consider the Ginzburg-Landau system of
equations for $\mathbf{A}$ and $\Psi$
\cite{tinkham}
\begin{eqnarray}
{1 \over 2m} ({\hbar \over i}\nabla  -2 e \mathbf{A})^2 \Psi -\alpha \Psi + \beta \Psi |\Psi|^2=0,\label{gl1} \\
\mathbf{J} = Im \left (  \Psi^* ({\hbar \over i}\nabla - 2 e \mathbf{A}) \Psi \right )~ , \label{gl2}
\end{eqnarray}
where $e$ is the charge of the electron and $m$ its effective mass.
We introduce the coherence length $\xi$, the equilibrium order parameter
$\psi_0^2$ and the London penetration depth $\lambda$ as
\be\label{xipsila}
\xi =\sqrt{ \hbar^2 \over 2 m \alpha},~~
\psi_0^2 = {\alpha \over  \beta},~~
\lambda=\sqrt{ m  \over 4 \mu_0 e^2 \psi_0^2}~.\ee
Substituting these quantities in the Ginzburg-Landau equations we get
\be\label{gl1a}
- \left (  \nabla -i{2e \over \hbar} \mathbf{A} \right )^2 \Psi
-{\Psi \over \xi^2}
+{4 \mu_0 e^2 \over m} {\lambda^2  \over \xi^2} \Psi |\Psi|^2=0~~. \ee
The equation for the current becomes
$$\mathbf{J} = -{1 \over \mu_0 } {1 \over \lambda^2 } \mathbf{A}
+ {2 e \hbar \over m} Im \left ( \Psi^* \nabla \Psi \right )
$$
Collecting all the terms of $\mathbf{J}$ and substituting into
Maxwell's equation, we obtain the more general model
\be\label{lmaxa3}
\Delta \mathbf{A} = {\mathbf{A}\over \lambda^2} -{2 e \hbar \over m}
Im \left ( \Psi^* \nabla \Psi \right ),\ee
containing the vortex contribution. The comparison with the
experiments presented below shows that (\ref{lmaxa}) provides a
good description of the fields around MgB$_2$ disks at 20K and
for applied fields below 0.4 T.
For these type II superconductors where $\kappa = \lambda/\xi >> 1$,
the decay distance of the order parameter
$\xi$ is much smaller than the decay distance of the field,
$\lambda$. Then the size of the vortices is small and the
correction on the right hand side of (\ref{lmaxa3})
due to $\Psi$ can be ignored in a first approximation.

\begin{figure}[H]
\centerline{
\resizebox{15 cm}{8 cm} {\includegraphics{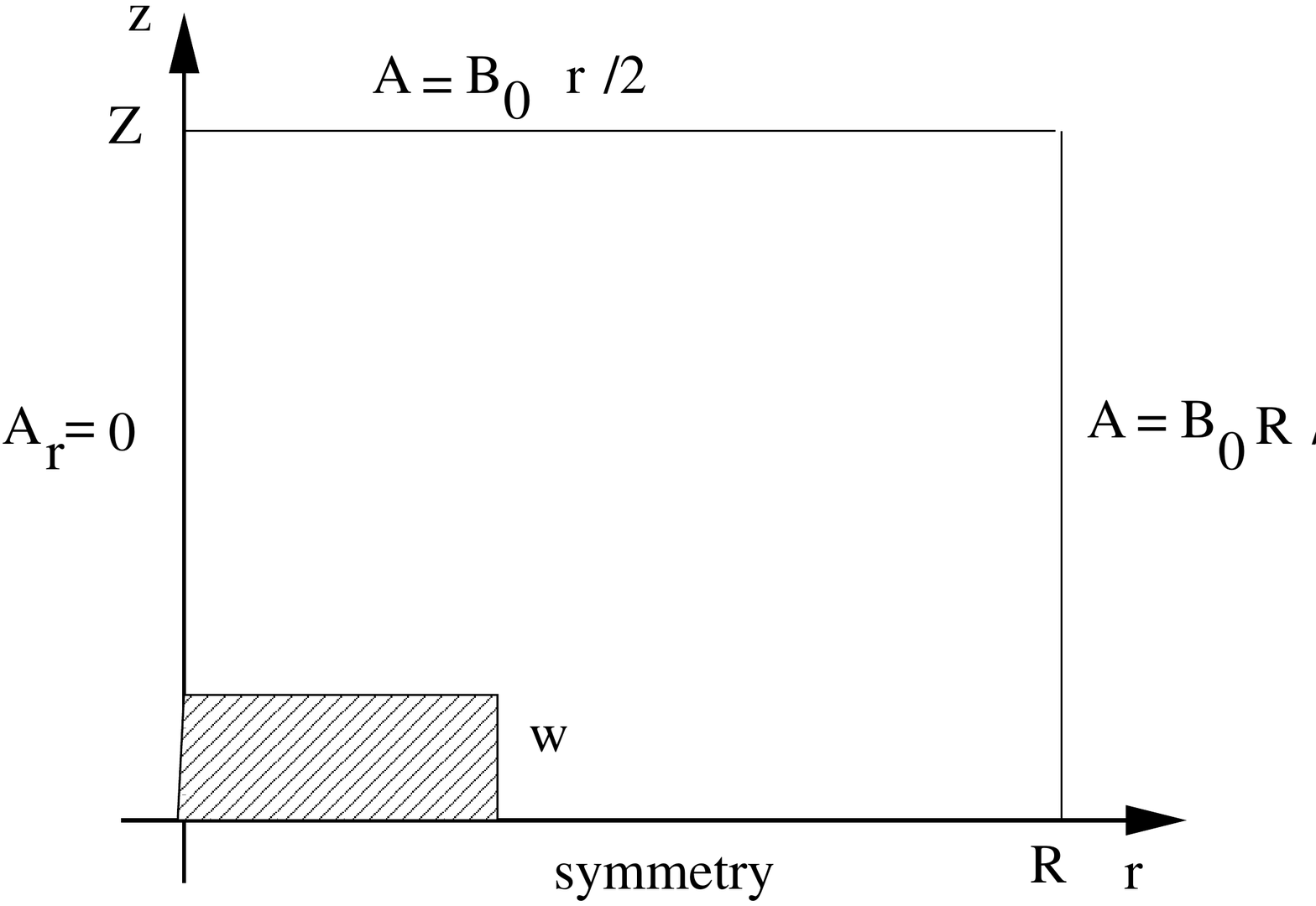} }
}
\caption{Computational domain for the solution of equation (\ref{lmaxa2})
in the $(r,z)$ plane. The boundary conditions are indicated.}
\label{scheme}
\end{figure}

In the experiment we used disk-shaped MgB$_2$ samples placed on the axis of a solenoid producing a constant field $B_0$ as in \cite{laura11}. Therefore, in order to reproduce the experimental results, in the model we can assume a cylindrical symmetry for the magnetic field $\mathbf{B}$.
Then the vector potential has only one component
$$\mathbf{A} = A {\vec \theta},$$
and is such that
\be\label{bofa}\mathbf{B} = \nabla\times \mathbf{A}
= -A_z \mathbf{r} + {1\over r}(rA)_r \mathbf{z}.\ee
There $\mathbf{r},\mathbf{z}$ are the unit vectors along the
$r$ and $z$ directions, respectively, and
the underscores represent partial derivatives.
Since $\mathbf{A}$ can be considered as a scalar, the
equation (\ref{lmaxa}) reduces to
\be\label{lmaxa2}
\Delta {A} = {1 \over \lambda^2} I(r,z) {A}.\ee
This equation for $A$ needs to be integrated in the $(r,z)$ plane.
The computational domain is shown in Fig. \ref{scheme} for the case of
a disk of thickness $2 w$, a sample that is symmetric with respect to
the plane $z=0$. The boundary conditions are indicated on Fig. \ref{scheme}.
For $r=0$ the magnetic field is along $z$ so $A_z=0$. At a large
distance from the sample,
the field is assumed constant, equal to $B_0$ and parallel
to $z$. The boundary condition is then $A=B_0 R/2$ where $R$ is
the edge of the solenoid generating the field.
To summarize we have the following
\begin{eqnarray}
z=0, ~~A ~~{\rm symmetric},\label{bc1}\\
z =Z >>0 , ~~A= {B_0 r \over 2},\label{bc2} \\
r=0, ~~A_z=0,\label{bc3}\\
r=R, ~~A={B_0 R \over 2},\label{bc4}
\end{eqnarray}
The only approximation is that we assume the field to be equal to $B_0$ for
large $z=Z$. Typically we took $Z=100w$ and made
sure that the results do not depend on this value. Of course if the
sample is not symmetric with respect to $z$ we need to consider the
two boundaries $z=\pm Z$.

\begin{figure} [H]
\resizebox{12 cm}{8 cm}{\includegraphics{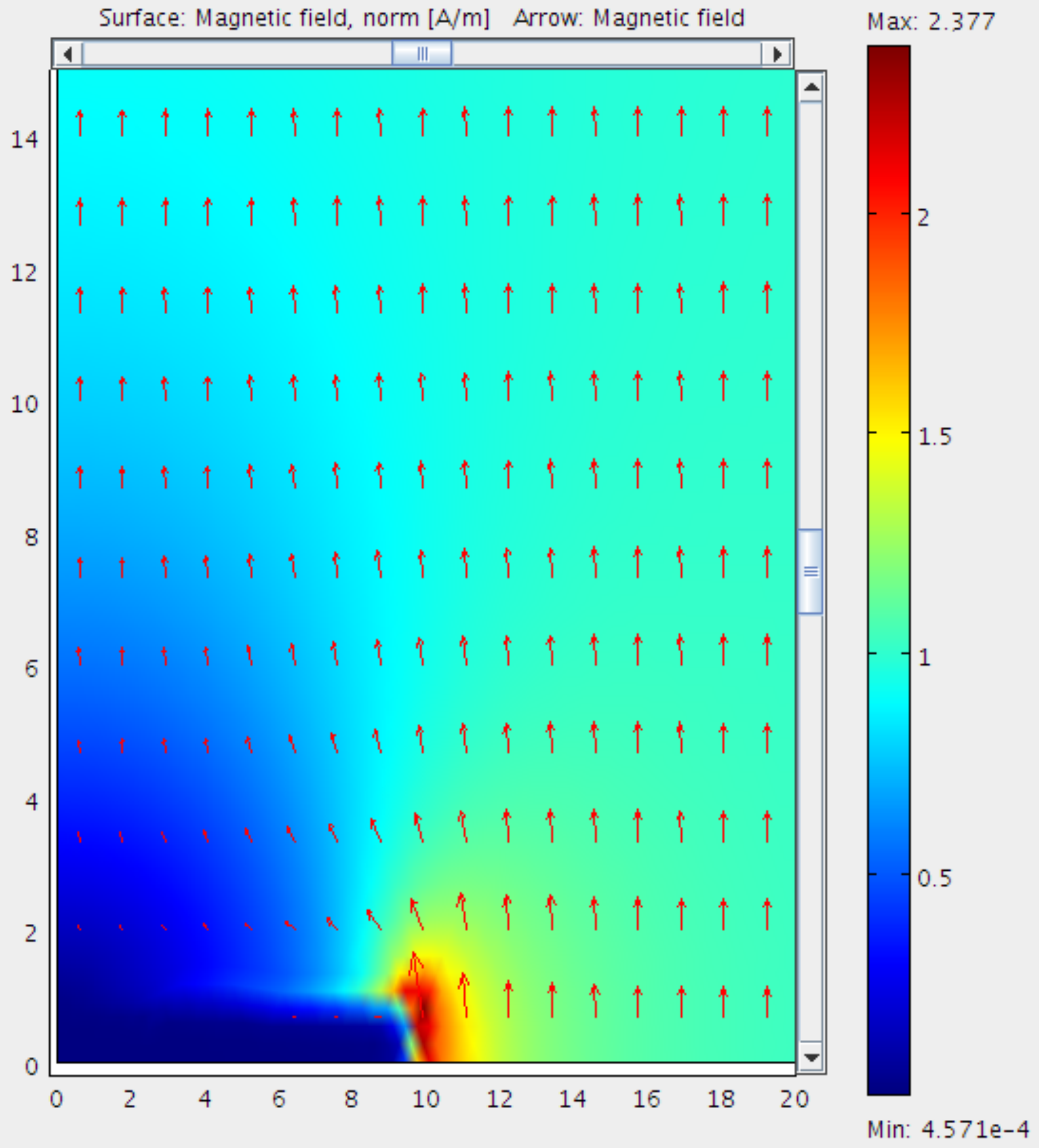} }
\caption{Numerical integration of the Maxwell/London equation (\ref{lmaxa2})
showing $B$ as a vector field for the disk geometry $D_1$ with $|B_0|=1$.}
\label{map_b}
\end{figure}

We present results obtained by solving equation (\ref{lmaxa2})
using the finite element software Comsol \cite{comsol}. As stressed above,
the problem
is linear so that $A$ can be scaled arbitrarily. Also the unit of
length has been chosen as $mm$ for commodity. Then the dimensions
of the sample and the London penetration depth are all given in mm.
The London penetration depth we have
chosen at 20K is $\lambda=1.6 10^{-4} mm$. It is in the range of
the measurements reported in  \cite{lambda}. Concerning the boundary
conditions in Fig. \ref{scheme} we stress that the
position of the boundary $z=Z=40$ is arbitrary. It corresponds to a value
for which the screening field has decayed enough so that $B=B_0$.
Fig. \ref{map_b} presents a typical result
of the magnetic field $B$ for an applied field $B_0=1$ for
the disk geometry $D_1$ (see Table 1 below). Since
the problem is linear, the magnitude of $B_0$ can be chosen arbitrarily.
$B$ ranges from 0 to 2.4 and is near zero in the superconductor.
The curvature of the flux-lines outside the superconductor
reduces the induction field near the upper surface of the
superconductor.
This effect is reinforced as the radius of the disks increases.
We emphasize the field reinforcement at the boundary $r=9.75$ mm of the disk.
In fact the field at the interface is singular in this model because
of the jump in $\nabla A$.
This model allows to calculate the magnetic induction field distribution 
everywhere
around a superconducting sample. It avoids the complications arising
from the computation of the demagnetizing field, even when the
external field is inhomogeneous.
However, the main advantage of this model is the possibility to solve
the inverse problem of designing a magnetic shield using as a
starting constraint the external applied field and
the geometry of the region to be shielded.
Of course this approach can be rigorously applied
when the flux density penetration
in the sample can be disregarded. However,
for initial vortex penetration at the surfaces in large
geometries as expected for shielding applications,
the outside field will not be radically different from
the one calculated
using our approach. This is due to geometric effects resulting from the
Laplacian equation or, in other words, to the demagnetizing energy
of the bulk Meissner state.
We will come back to this point below.

\section{Samples fabrication and characterization technique}

Three disk-shaped MgB$_2$ samples were fabricated by non-conventional
Spark Plasma Sintering (SPS) \cite{FCT}.
Their dimensions are reported in Table I.
\begin{table} \label{tab}
\begin{tabular}
{|l | c | c | r |}
  \hline
disk            &  diameter (mm)   &  thickness (mm) \\ \hline
$D_1$               & 19.5             &1.9 \\ \hline
$D_2$               &  14.8              &1.75 \\ \hline
$D_3$               & 8.1             &1.8 \\ \hline
 \end{tabular}
\caption{Dimensions of the three disks analyzed.}
\end{table}
The samples were fabricated by pouring the commercially available
MgB$_2$ powder \cite{Starck}
into a graphite mould, that was placed into the working chamber
that was evacuated down to a pressure of 1 mbar.
A pulsed electric current (2000 A, 4 V) was passed through the sample
while the temperature was raised to 1200$^\circ$C in 7 min.
The samples were kept at this temperature for 5 min under a
50 MPa uniaxial pressure. Finally, they were cooled down to
room temperature in 8 min. The disks obtained were rectified
by mirror polishing using pure ethanol as a lubricant. The relative
density of the samples was more than 98 \%
of the theoretical value, their Vickers hardness was 1050 MPa and
their critical temperature was T$_c$ = 37 K.

The measures were carried out at a temperature of 20 K in a
uniform dc magnetic field up to 1.5 T. These fields are applied
in the $z$ direction, perpendicularly to the sample surface.
They are generated by a superconducting cryogen free coil
coaxial to the samples. The samples were mounted on the top
of the second cooling stage of a cryocooler with an interposed
0.125 mm thick indium sheet in order to guarantee a good thermal
contact to avoid thermo-magnetic instabilities causing
flux jumps \cite{aj04,glgg07}. These would strongly modify the
shielding capability of the sample and, in the worst case, could
create cracks and irreversible damage.
A schematic view of the experimental set-up is reported in \cite{gmggl11}.

\begin{figure} 
\centerline{
\resizebox{12 cm}{4 cm}{\includegraphics{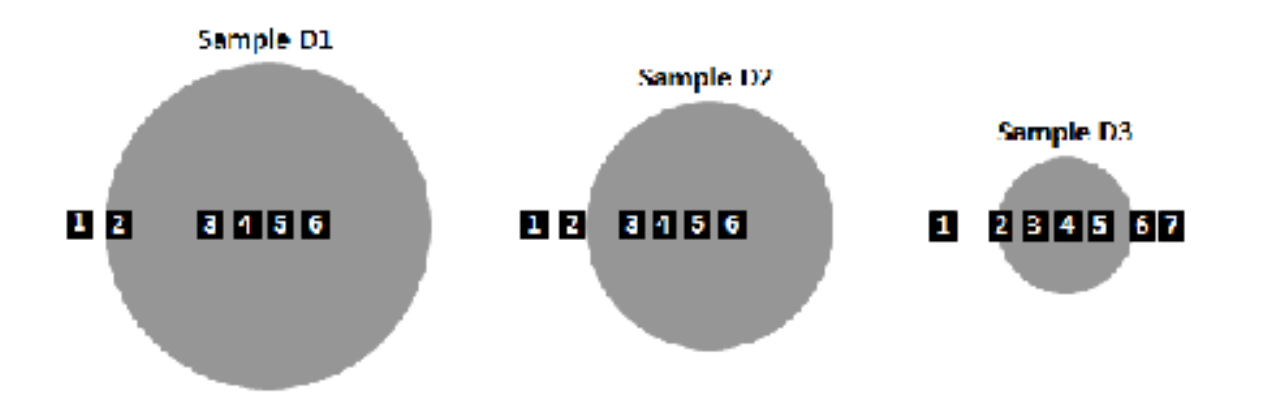} }
 }
\caption{Hall probe arrays for measuring the $z$ component of the
magnetic field for the three samples of table I.}
\label{sonde}
\end{figure}

The $z$ component of the magnetic induction, i.e. the component parallel
to the applied magnetic field direction, was measured with a GaAs Hall
probe array mounted on the bottom surface of a custom-designed
motor-driven stage, able to be moved along the sample axis with
a spatial resolution of 1$\mu$m. Each probe has a disk-shaped active
area with a diameter of 300 $\mu$m and an average sensitivity of 43.2 mV/T
for a bias current of 0.1 mA. The probes were aligned along the sample
diameter following the radial arrangement reported in Fig. \ref{sonde}.
The radial positions are detailed in table II for each sample.
\begin{table} \label{tab2}
\begin{tabular}
{|l | c | c | c | c | c | c | c | r |}
  \hline
disk    &  radius & d & d & d & d &  d   &  d & d \\ \hline
D1      & 9.75 &   -11.2 &  -9.8 &  -3.2  &  -1.2  &  0.8  &   2.8&  \\ \hline
D2      &  7.4 & -10.5&    -8.2&  -4.7 &   -2.7 &   -0.7 &   1.3  &  \\ \hline
D3      & 4.05 &-7.4 &   -3.9   & -1.9  &  0.1  &  2.1   &  4.6  &   6.4  \\ \hline
 \end{tabular}
\caption{Radial distance of the probes from the center for each sample. All the dimensions are in mm; the error in the position of each probe is about 0.2 mm.}
\end{table}

The sample temperature, the applied magnetic field, the Hall
probe positioning and the Hall voltage were controlled with a
Labview$^{TM}$ custom program.
The experiments were performed after a zero-field cooling. The magnetic
field was gradually increased up to a predetermined value. The
induction field profiles were recorded for different distances
$z$ above the sample.

\section{Analysis of the experimental data}

An important consequence of the
London approximation is that the model is linear so the
results should scale with the magnetic field $B_0$. We have tested
this scaling on the experimental data for the three disks analyzed.
The main result is that for all three samples the experimental curves
scale as $B/B_0$ as long as $B_0 <$ 0.4T. The scaling is perfect
up to 0.1T and above that value there are small differences
especially close to the center of the disk. It is worthwhile
to remember that the data refer
only to the $z$ component of the field $B$.

\begin{figure}
\centerline{ \resizebox{12 cm}{5 cm}{\includegraphics{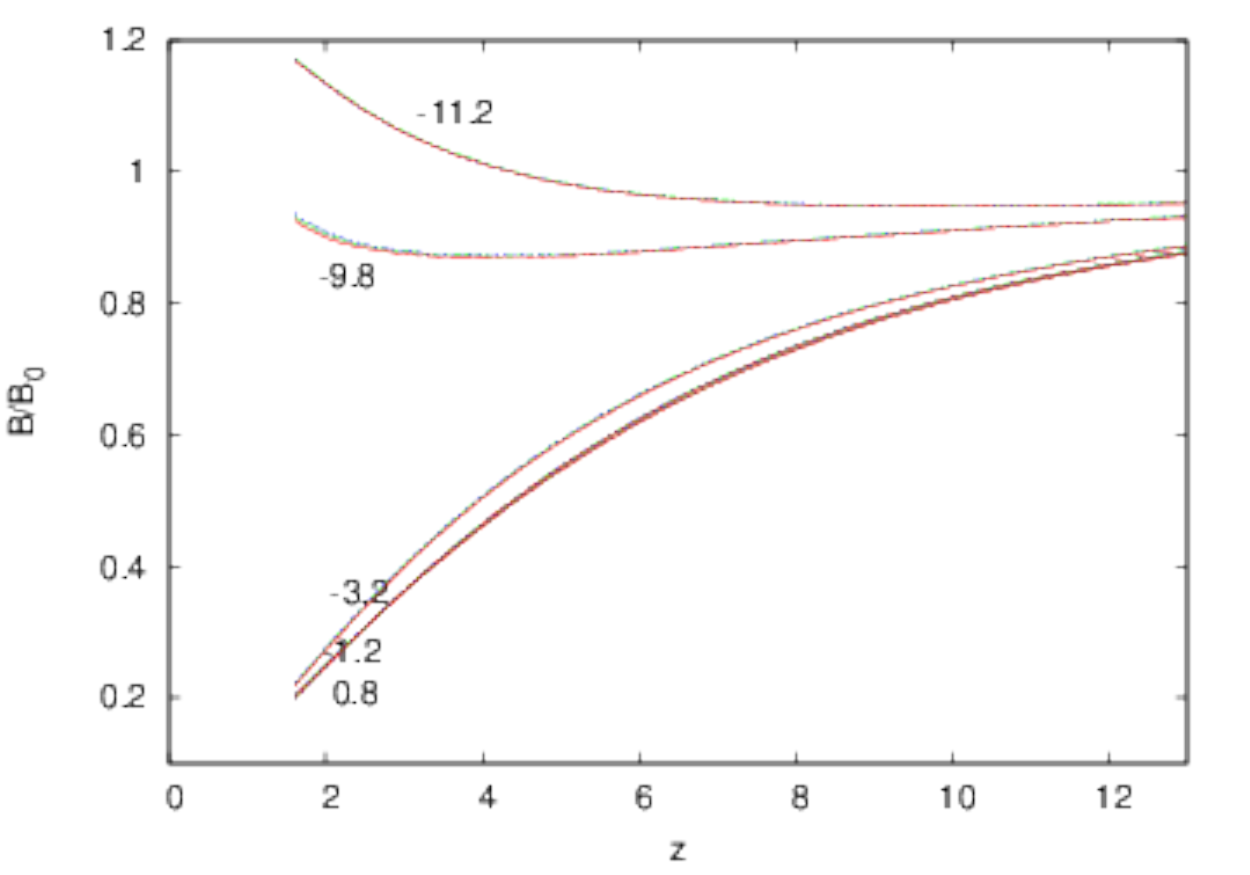} }}
\centerline{ \resizebox{12 cm}{5 cm}{\includegraphics{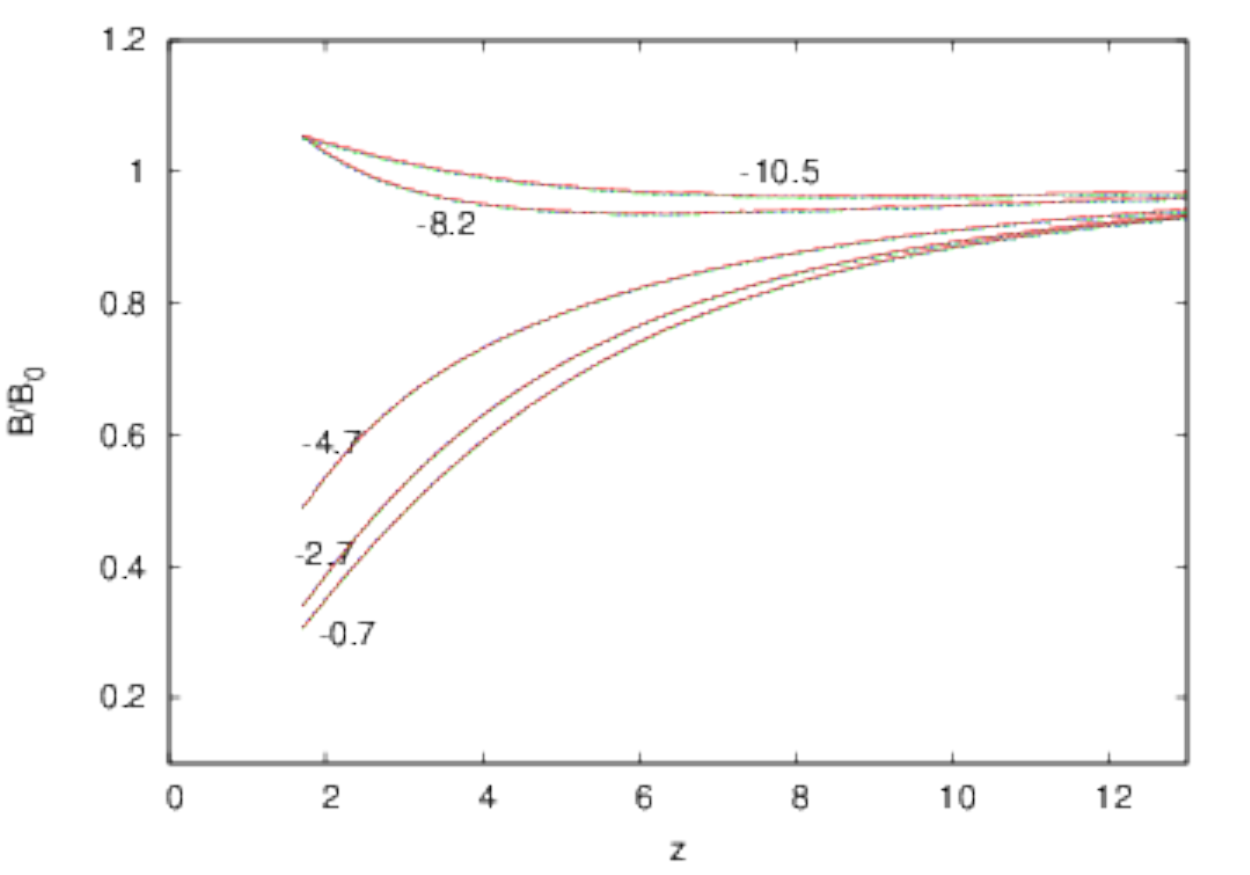} }}
\centerline{ \resizebox{12 cm}{5 cm}{\includegraphics{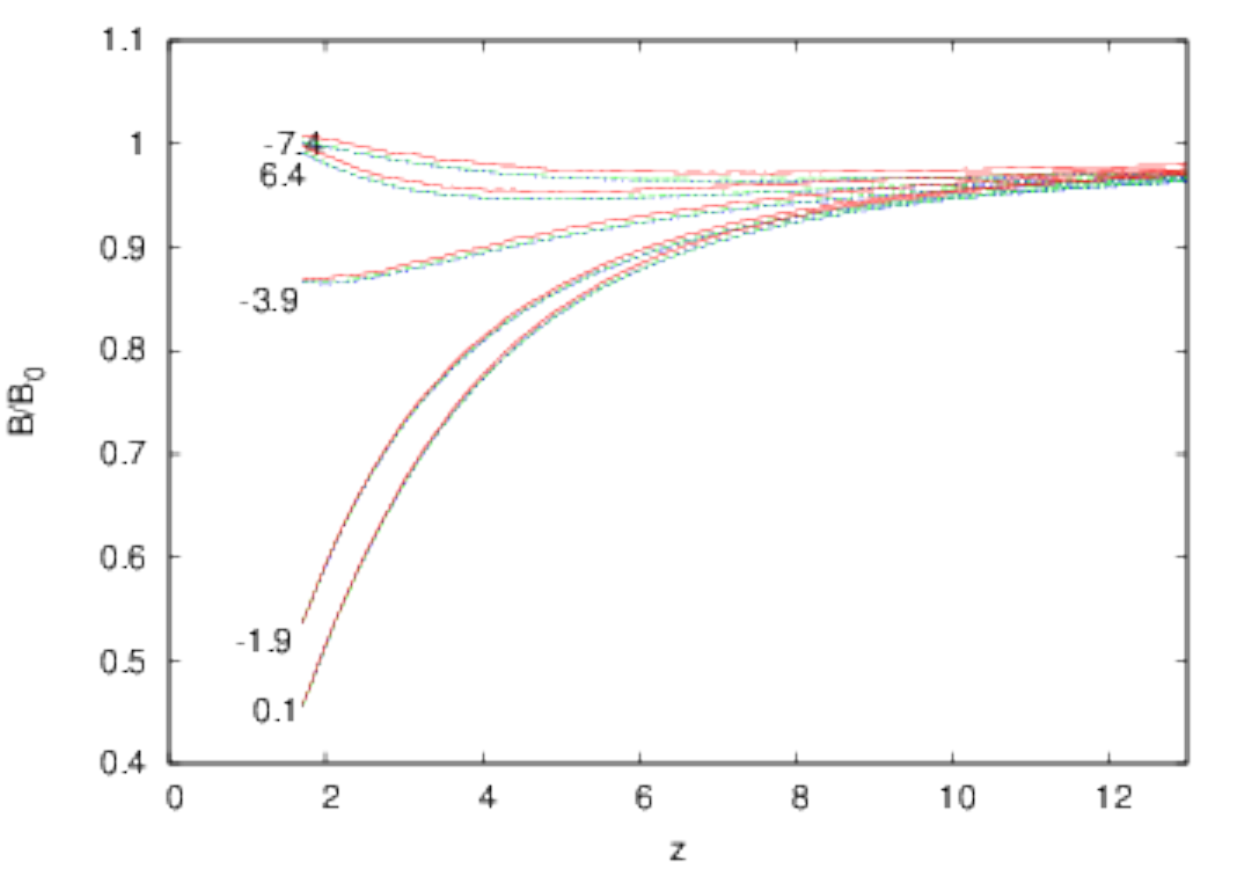} }}
\caption{
Plot of $B/B_0$ as a function of the distance above the superconductor $z$ for
$B_0$ = 0.04T (continuous line, red online), for $B_0$ = 0.07T
(long dashed line, green online) and $B_0$ = 0.1T (short dashed line,
blue online) and
for different radial positions. The three panels correspond
to the three samples $D_1,~D_2,~D_3$ from top to bottom.}
\label{f2}
\end{figure}

We first show the results for a small applied field $B_0 < $ 0.1T.
In Fig. \ref{f2} we present $B/B_0$ as a function of the distance $z$ from
the superconductor surface for
$B_0$ = 0.04T, 0.07T and 0.1T
at different radial positions for the three samples $D_1,~D_2,~D_3$. In
the following whenever we consider measurements $z$ will refer to
the distance above the superconductor.
As expected there is a very good scaling. The shielding effect
is maximum in correspondence to the sample center, increases
with sample diameter and decreases towards the edges.
The upper curvature of the field lines detected near the samples
by the outer Hall probes is due to the demagnetization effects
at the disk edges \cite{brandt98}. It is more pronounced
in the sample D$_1$ because it has the largest diameter.

\begin{figure}
\centerline{ \resizebox{12 cm}{5 cm}{\includegraphics{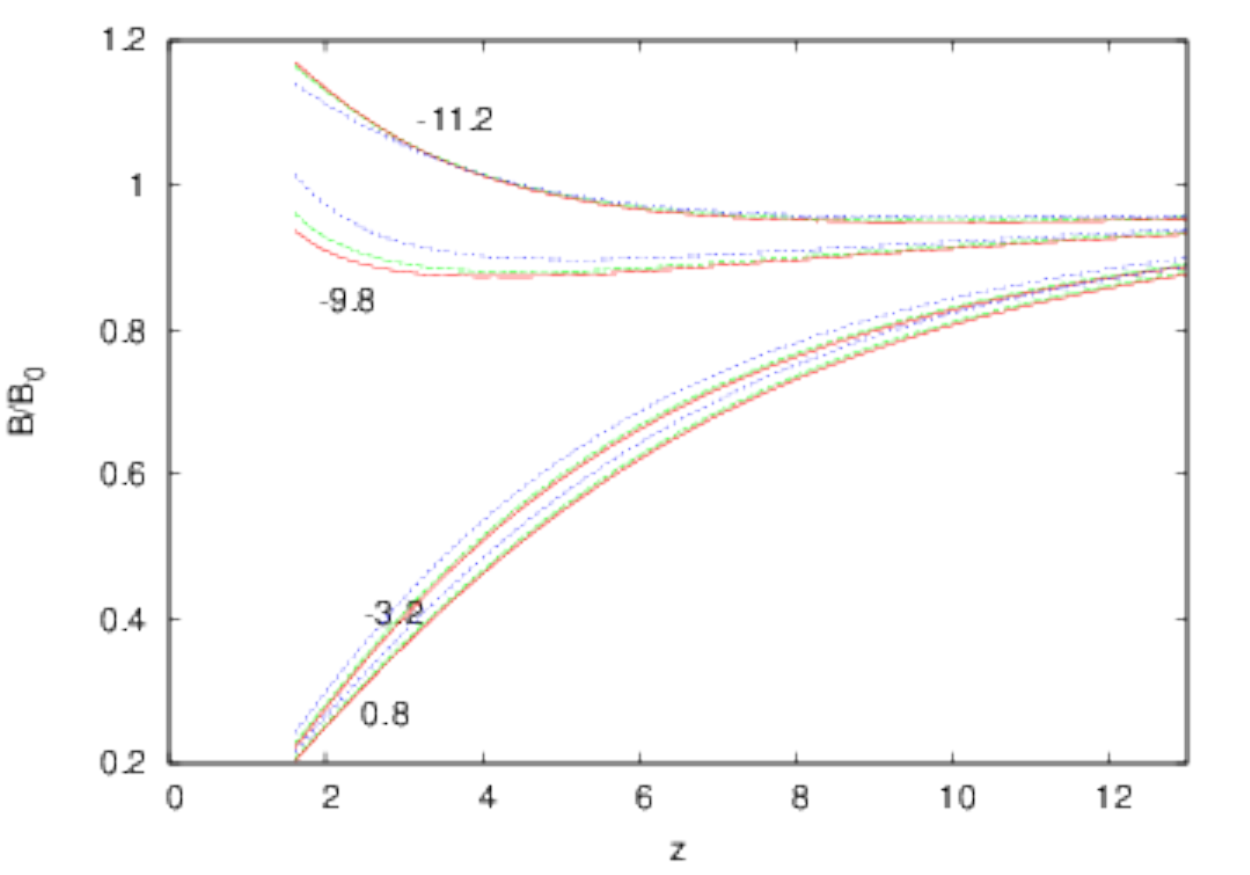} }}
\centerline{ \resizebox{12 cm}{5 cm}{\includegraphics{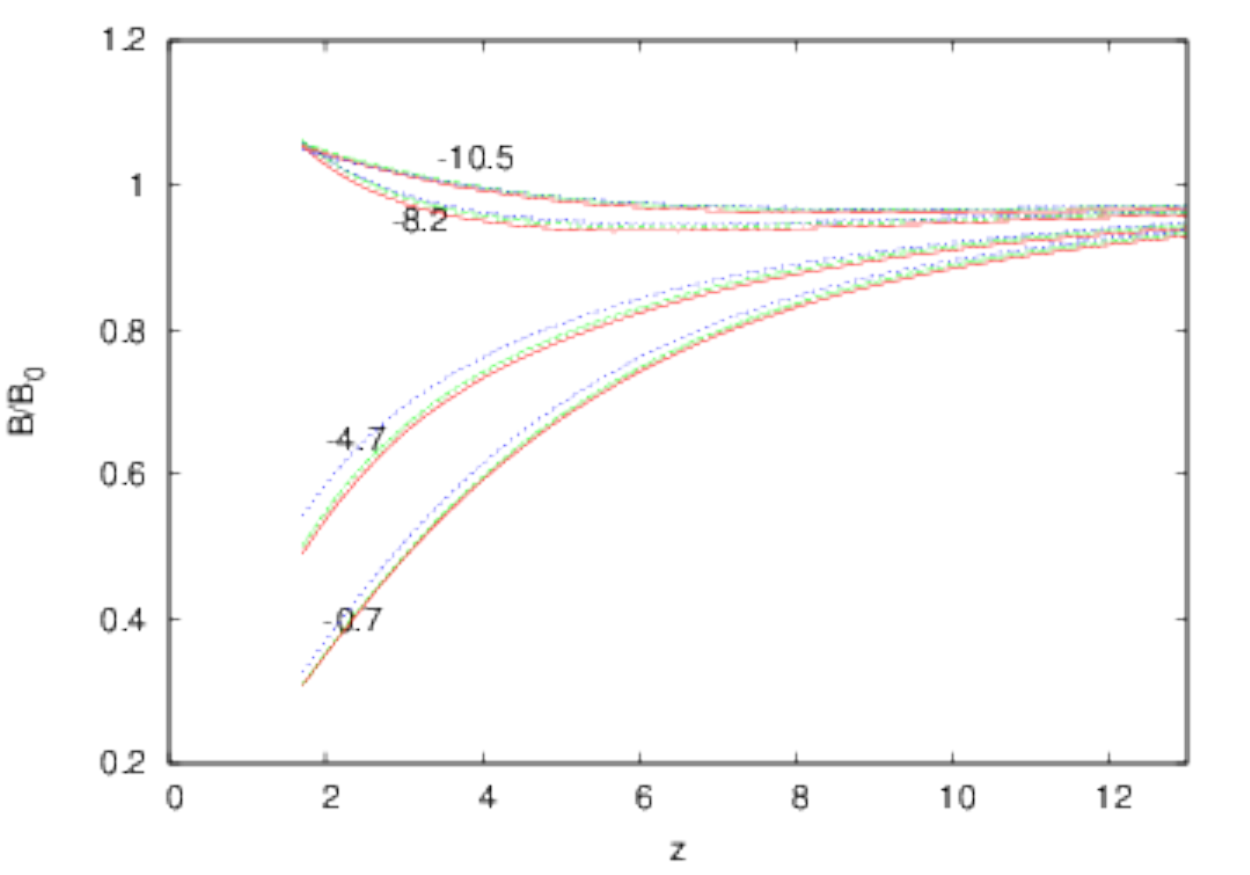} }}
\centerline{ \resizebox{12 cm}{5 cm}{\includegraphics{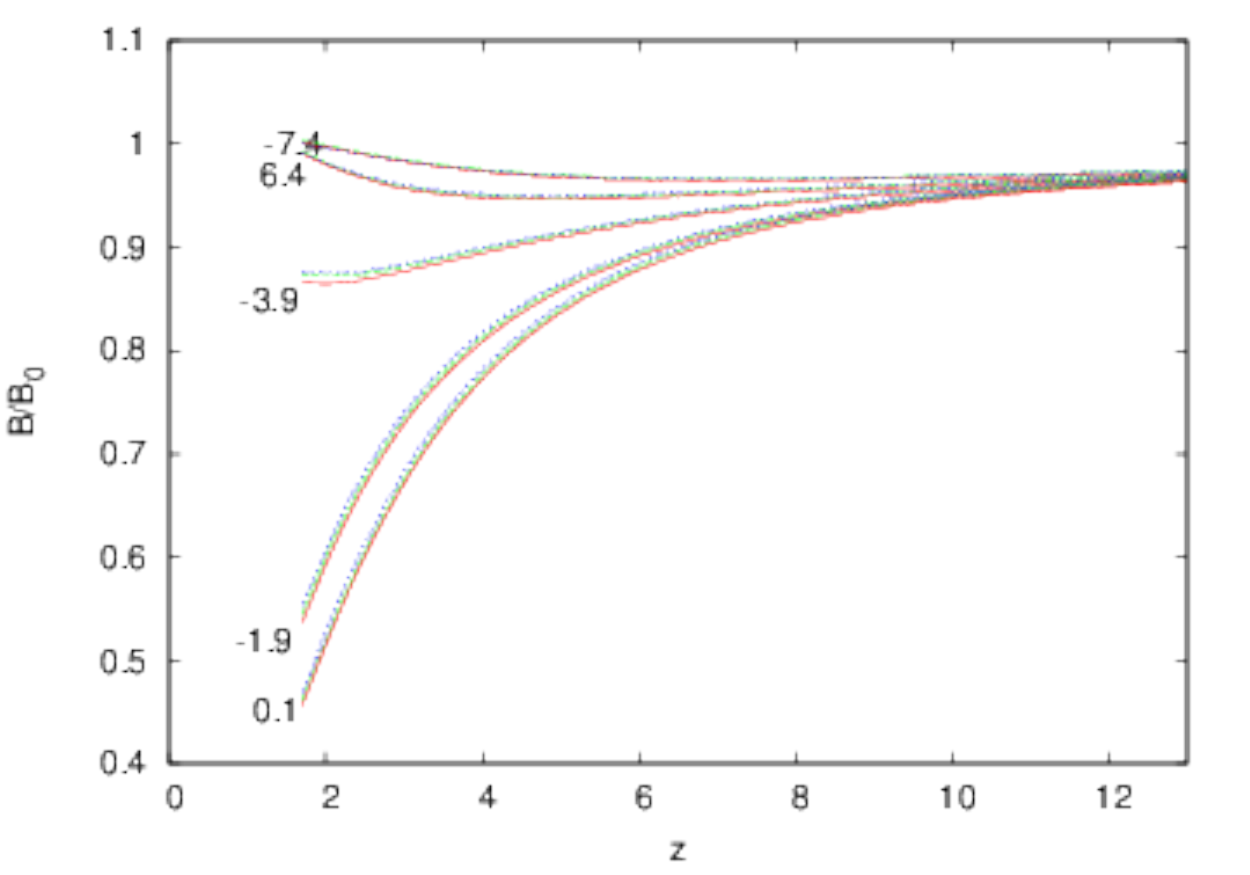} }}
\caption{
Plot of $B/B_0$ as a function of $z$ for
$B_0$ = 0.1T (continuous line, red online), 0.2T (long dashed line, green online) and 0.4T (short dashed line, blue online) and
for different radial positions. The three panels correspond
to the three samples $D_1,~D_2,~D_3$ from top to bottom.}
\label{f3}
\end{figure}

When the applied field is increased up to 0.4T  the scaling remains quite good even if some discrepancies start emerging. We show in Fig. \ref{f3} the ratio
$B/B_0$ as a function of $z$ for
$B_0$ = 0.1T, 0.2T and 0.4T. Again the three
panels correspond to the three samples.
These results confirm that a linear theory such as the Maxwell/London
equation can describe well the disk data for magnetic fields smaller
than 0.4T at 20K.

\begin{figure}
\centerline{
\resizebox{12 cm}{15 cm}{\includegraphics{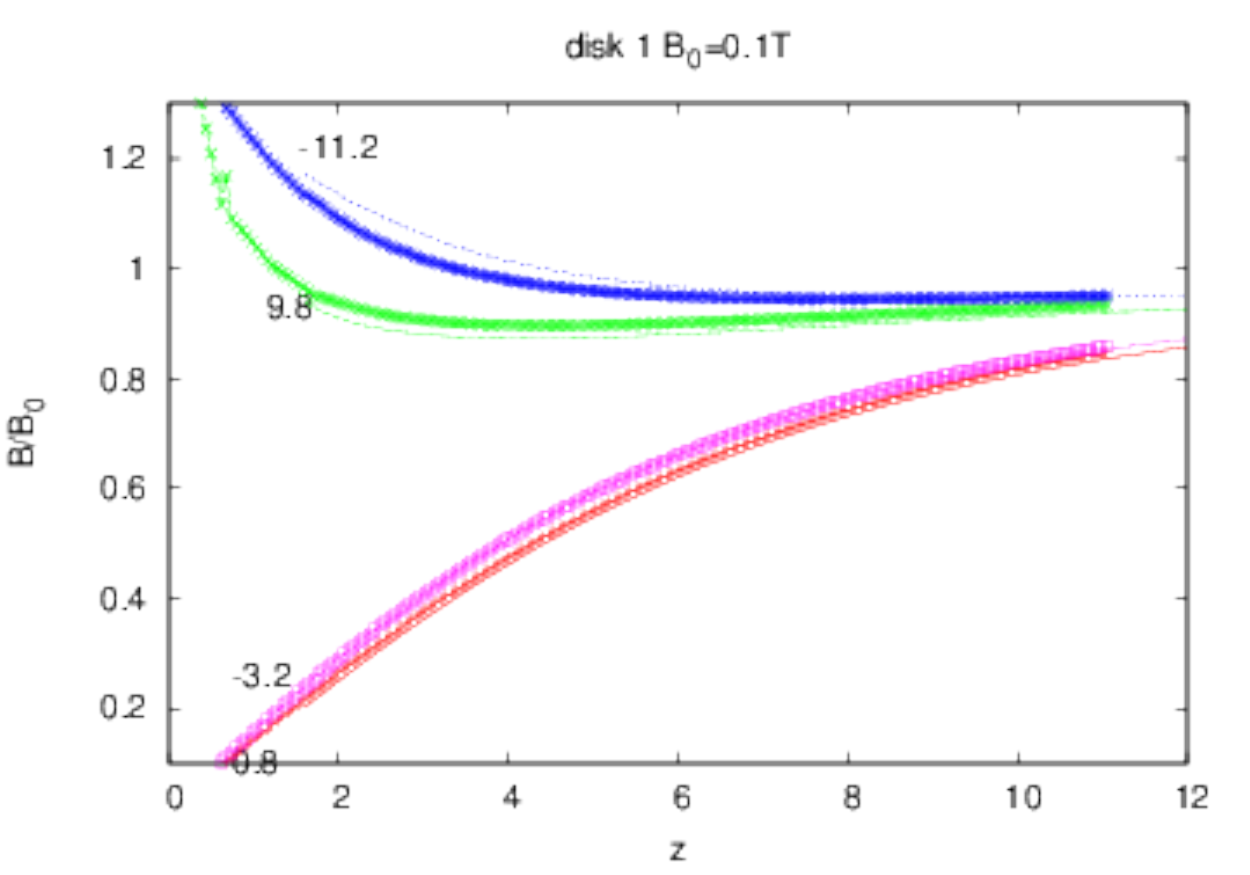} }
}
\caption{Comparison between the ratio $B/B_0$ calculated
and measured for sample D$_1$ as a function of the distance $z$
above the superconductor for different $r$. The measurement 
temperature was T=20K and $B_0$ = 0.1T.
The experimental data are shown with lines and the numerical values
are plotted with symbols. The different values of $r$ are
reported in the figure.}
\label{comp_d1}
\end{figure}

\begin{figure}
\centerline{
\resizebox{12 cm}{15 cm}{\includegraphics{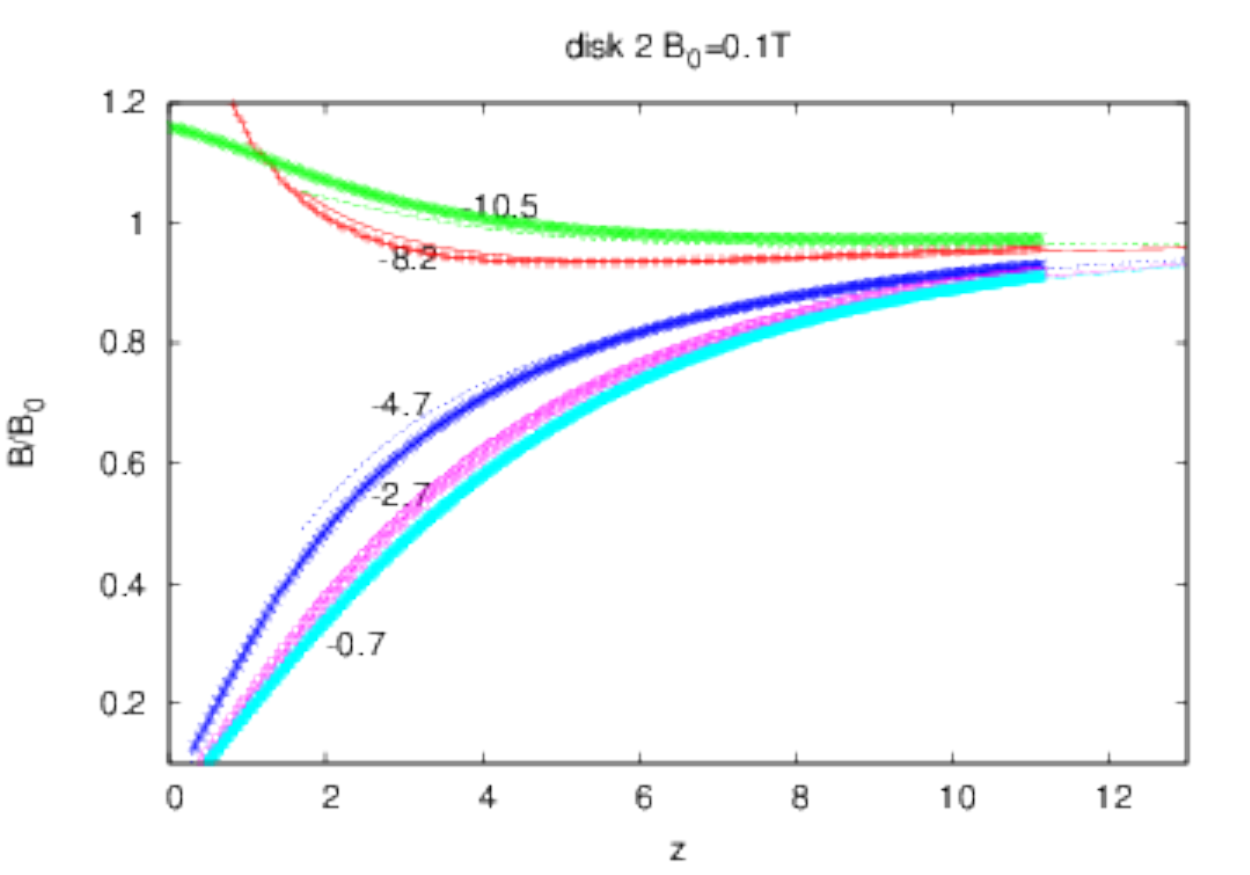} }
}
\caption{Comparison between the ratio $B/B_0$ calculated and
measured for sample D$_2$ as a function of $z$ and for different $r$.
The other parameters are the same as in Fig. \ref{comp_d1}.}
\label{comp_d2}
\end{figure}

\begin{figure}
\centerline{
\resizebox{12 cm}{15 cm}{\includegraphics{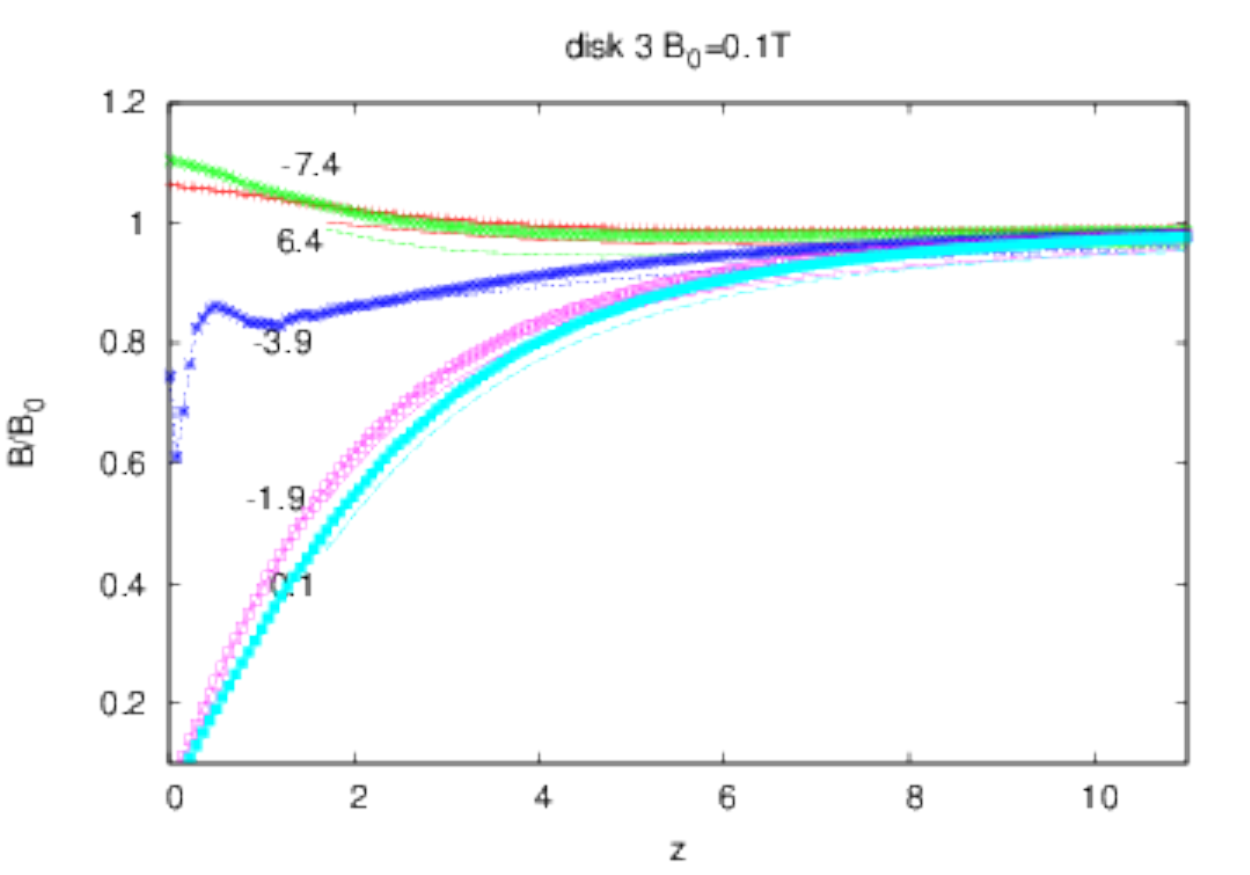} }
}
\caption{Comparison between the ratio $B/B_0$ calculated and
measured for sample D$_3$ as a function of $z$ and for different $r$.
The other parameters are the same as in Fig. \ref{comp_d1}.}
\label{comp_d3}
\end{figure}

We now use these values of $B_0$ to compare the solution of the
London equation (\ref{lmaxa}) with the experimental
data. The results are shown in
Fig. \ref{comp_d1} for the disk sample $D_1$. There the experimental data
are shown as lines for clarity. Only the value $B_0$ = 0.1T is presented
since we have a good scaling $B/B_0$
as shown previously. The agreement is good for small and large $r$ values.
For the disks 2 and 3, we observe a similar trend  as shown in
Figs. \ref{comp_d2} and \ref{comp_d3}.

\begin{figure}
\centerline{\resizebox{12 cm}{5 cm}{\includegraphics{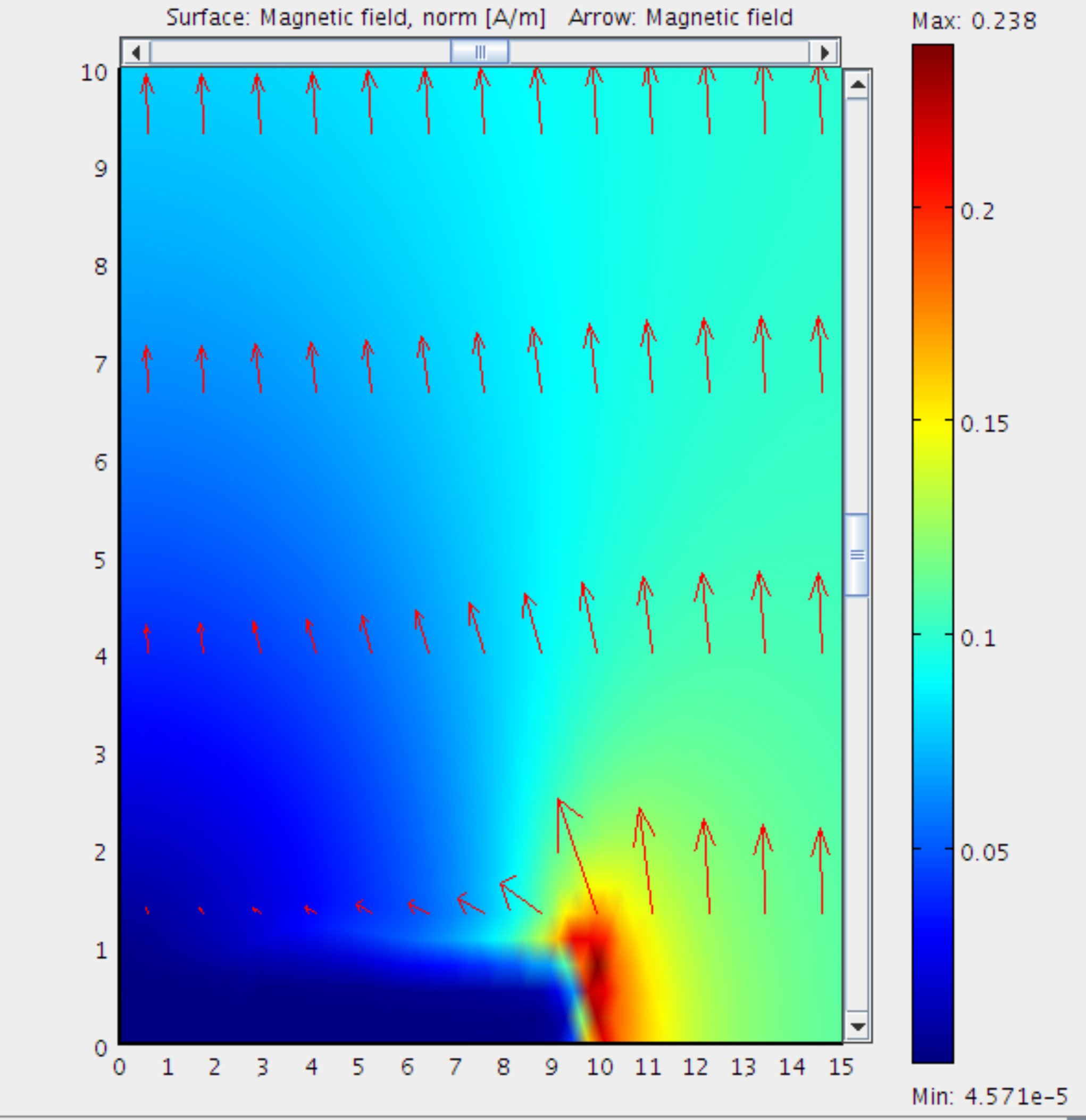} }}
\centerline{\resizebox{12 cm}{5 cm}{\includegraphics{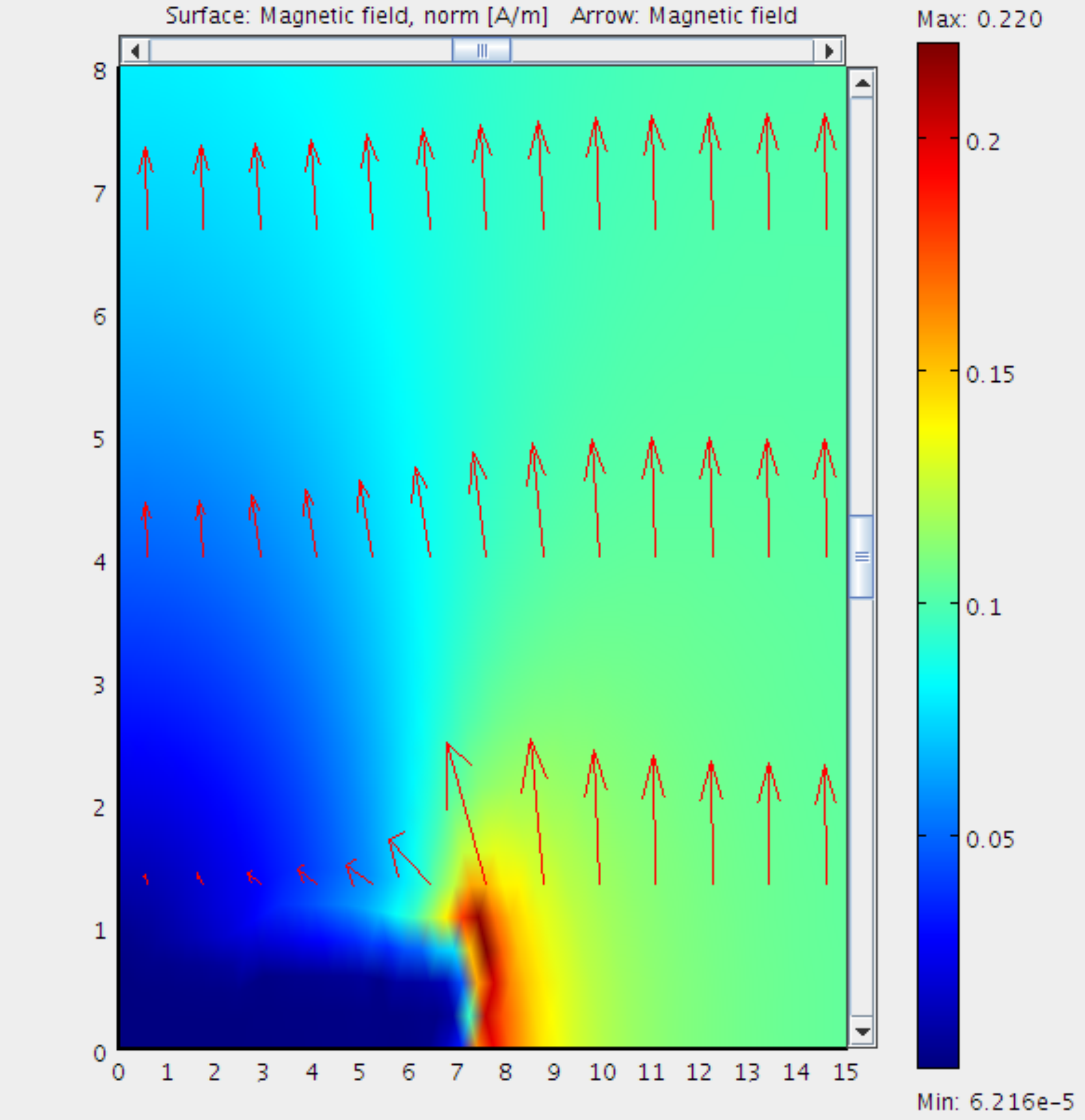} }}
\centerline{\resizebox{12 cm}{5 cm}{\includegraphics{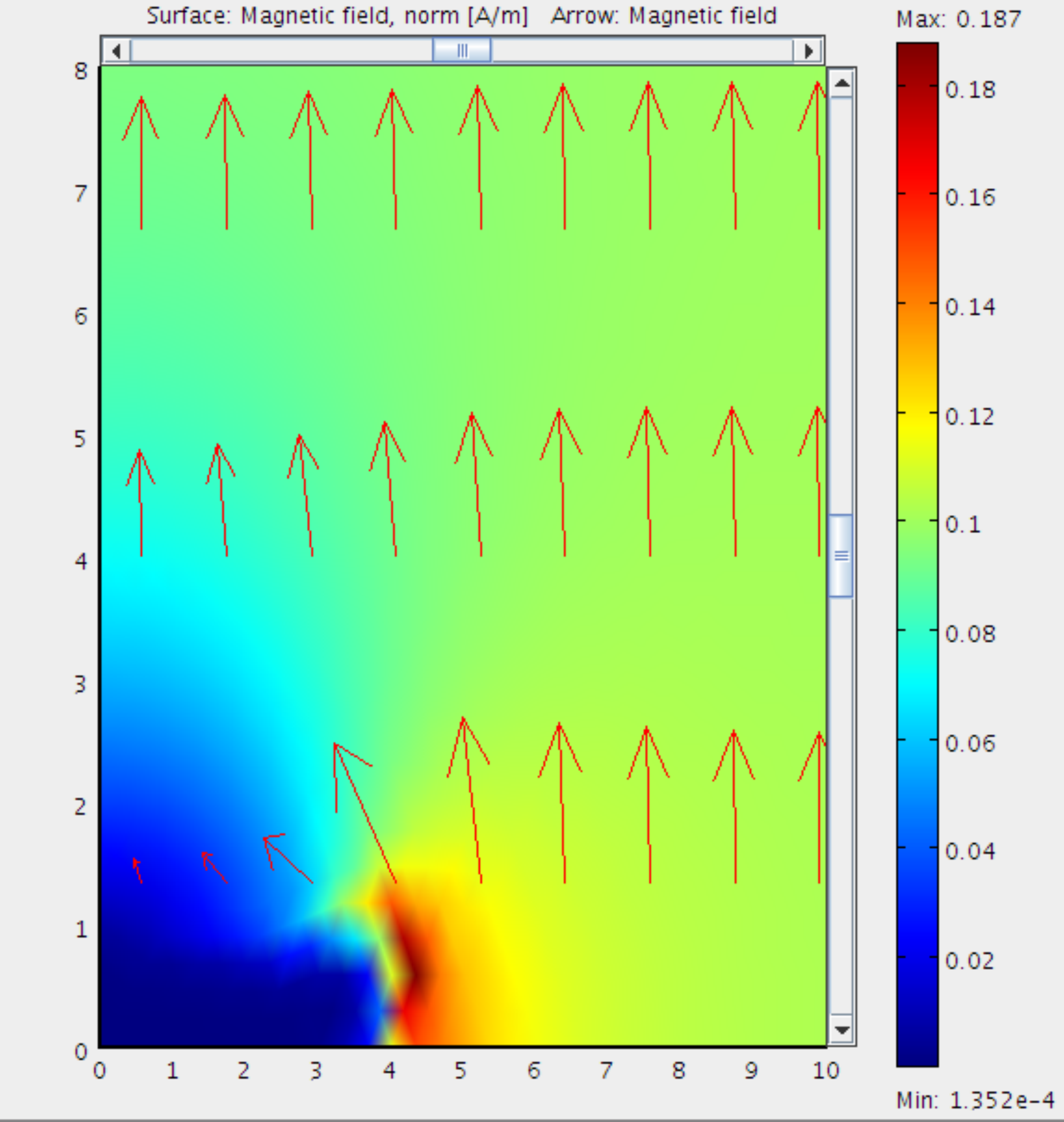} }}
\caption{Blow-up of the region close to the samples showing how screening
is enabled.
The magnetic induction field $\mathbf{B}$ is shown with arrows and its
magnitude is in the colour code.
The three samples are $D_1$ (top), $D_2$ (middle) and $D_3$
(bottom). The maximum of $|\mathbf{B}|$ in dark red (online) is
0.24 for $D_1$, 0.22 for $D_2$ and  0.19 for $D_3$. The $z$ range is
$z<10$ (top panel) and $z<8$ (middle and bottom panels). The applied
field is $B_0=0.1$.
}
\label{blow_up}
\end{figure}

When the field $B_0$ is increased, vortices penetrate the sample
so that the phase cannot be considered as uniform. Then
the vector potential $A$ does not depend linearly on the applied field $B_0$.
This effect is stronger for the small sample $D_3$ because it does
not screen the field as well as $D_1$ does.
Fig. \ref{blow_up} shows the region close
to the samples for $D_1$ (top panel), $D_2$ (middle panel)
and $D_3$ (bottom panel).

The screening region where the field is close to zero is a triangle
$z<R_1/2   -0.3 r $ for the sample $D_1$,
$z< R_2/2  -0.43 r $ for the sample $D_2$
and $z<R_3/2  -0.5 r $ for the sample $D_3$.
This screened region by disk $D_1$ is twice as large as the one
screened by disk $D_3$.
This is a geometric effect that depends on the sample dimensions only.
\begin{figure}
\centerline{\resizebox{12 cm}{5 cm}{\includegraphics{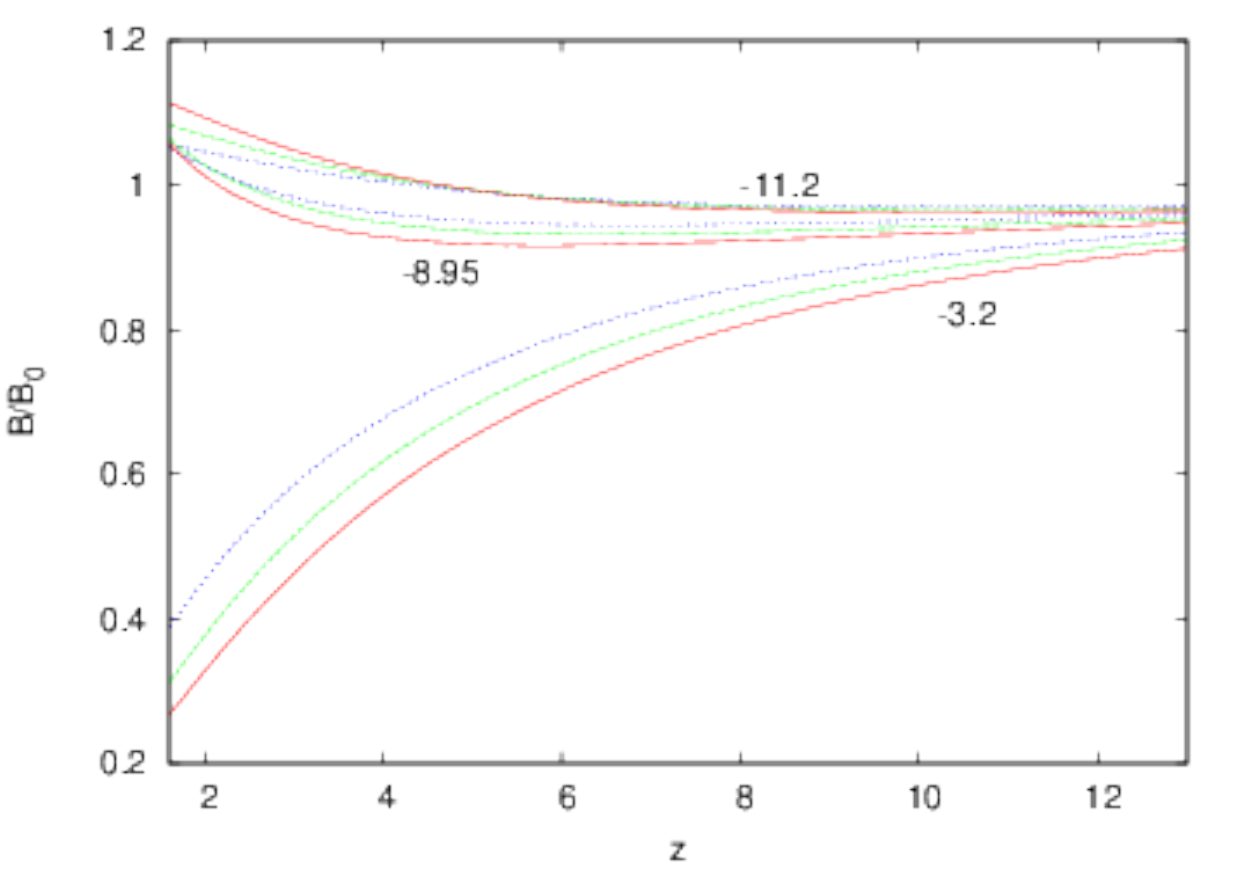} }}
\centerline{\resizebox{12 cm}{5 cm}{\includegraphics{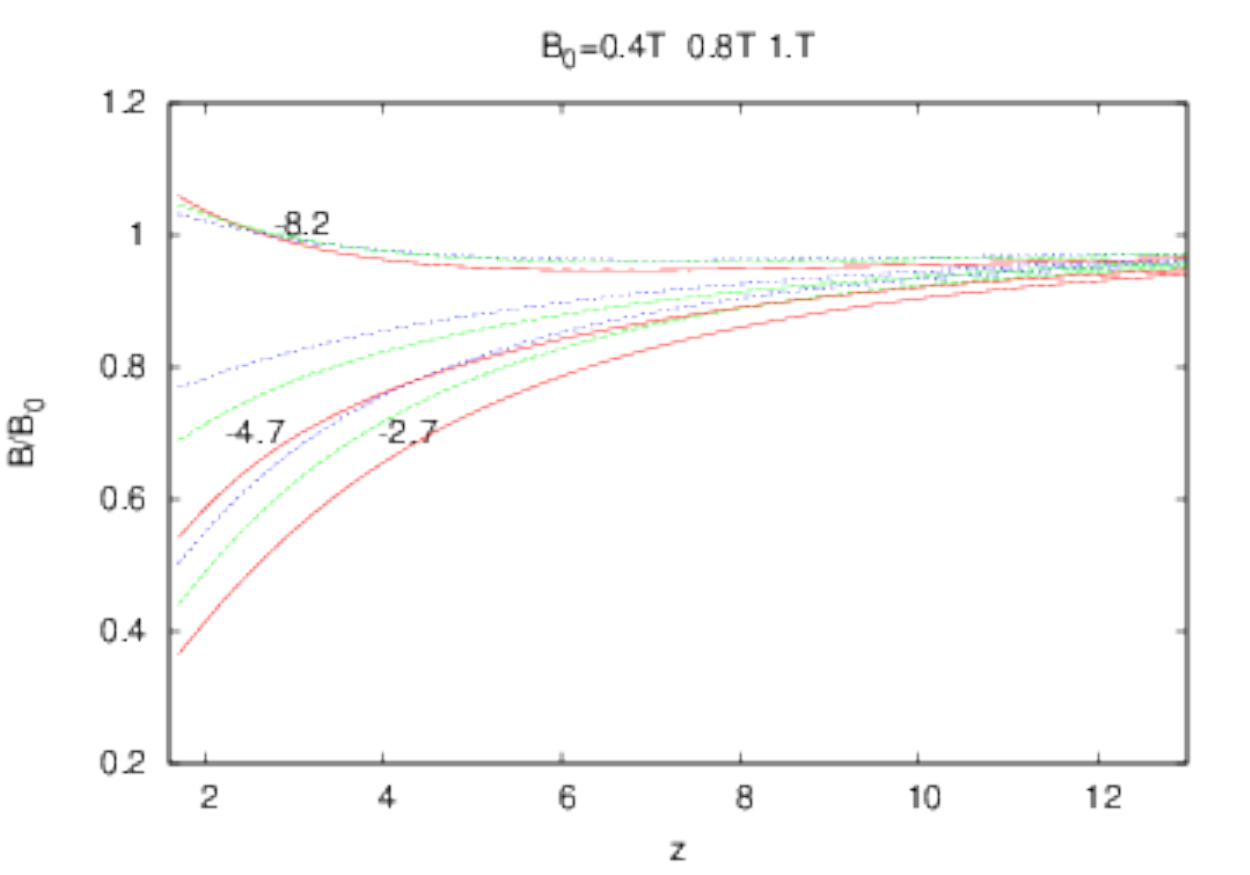} }}
\centerline{\resizebox{12 cm}{5 cm}{\includegraphics{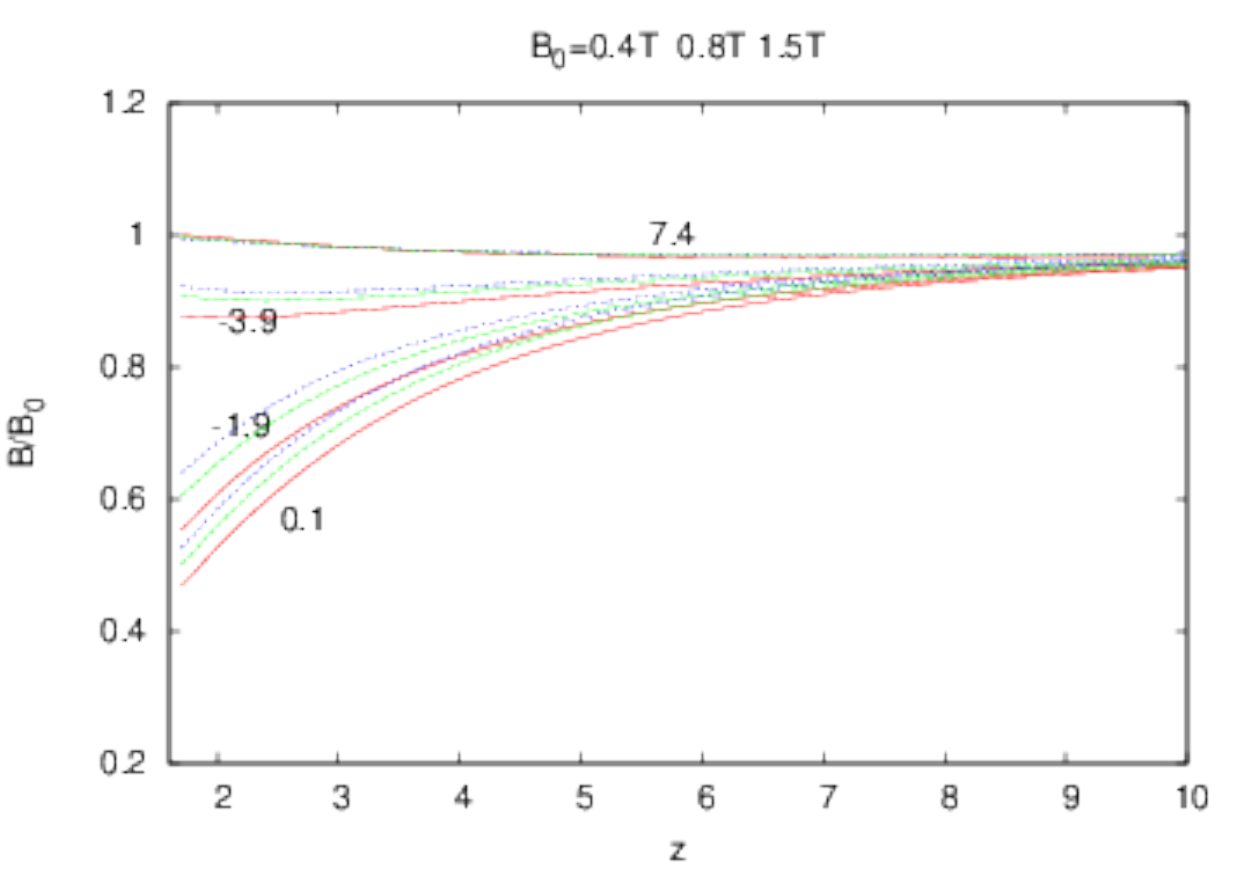} }}
\caption{
Ratios $B/B_0$ as a function of $z$ for
$B_0$=0.4T (continuous line, red on line), 0.8T (long dashed line, green on line) and 1T (short dashed line, blue on line) and
different radial positions. The three panels correspond
to samples $D_1,~D_2,~D_3$ from top to bottom.}
\label{d123d5}
\end{figure}
The $B/B_0$ ratios shown in Fig. \ref{d123d5} show that the scaling
is not so good for $B_0$ equal or higher than 0.4T, indicating that for such a large field the vortex
penetration inside the sample cannot no longer be disregarded. Nonlinearities
appear that can be taken into account in the model.

\section{Discussion and Conclusion}

\begin{figure}
\centerline{\resizebox{12 cm}{5 cm}{\includegraphics{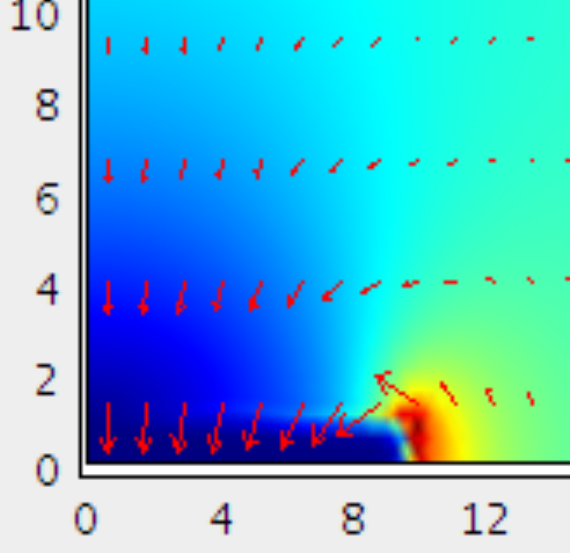} }}
\caption{ Screening field $\mathbf{B}-\mathbf{B_0}$ of disk $D_1$ shown as
arrows.
The total field, $\mathbf{B}$, is shown with the color code as in Fig. \ref{blow_up}.}
\label{screen_d1}
\end{figure}

From the results shown in the previous section, we see that the
Maxwell/London model (\ref{lmaxa}) is appropriate to
describe in a simple way the magnetic induction field distribution outside
disk-shaped MgB$_2$ samples at 20K for applied fields lower than 0.4T.
From the model, it is easy to compute the shielding field, $\mathbf{B}-\mathbf{B_0}$, generated by the disks. We show this field direction in Fig. \ref{screen_d1} with the arrows, while the modulus of the total field $\mathbf{B}$ is shown with the color code.
The superconductor induces a redistribution of the induction field
around itself.
In particular, notice the shielding effect above the upper surface of the sample
where the superconductor generates a field that is exactly opposed
to the applied field. There is also a strong field reinforcement
right outside the disk, for $z=0,r > R_1$ , where the shielding
field is aligned with the applied field.

\begin{figure}
\centerline{\resizebox{12 cm}{5 cm}{\includegraphics{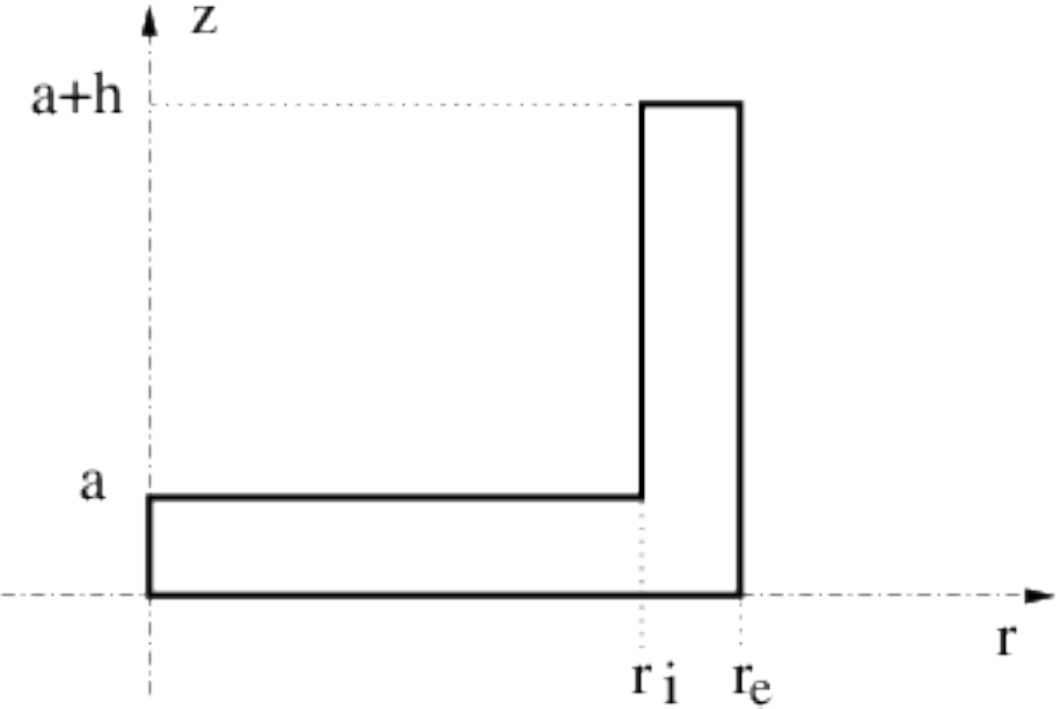} }}
\caption{ Schematic drawing of a magnetic screen in the form of a cup of
depth $h$.}
\label{cup}
\end{figure}

\begin{figure}
\centerline{\resizebox{12 cm}{5 cm}{\includegraphics{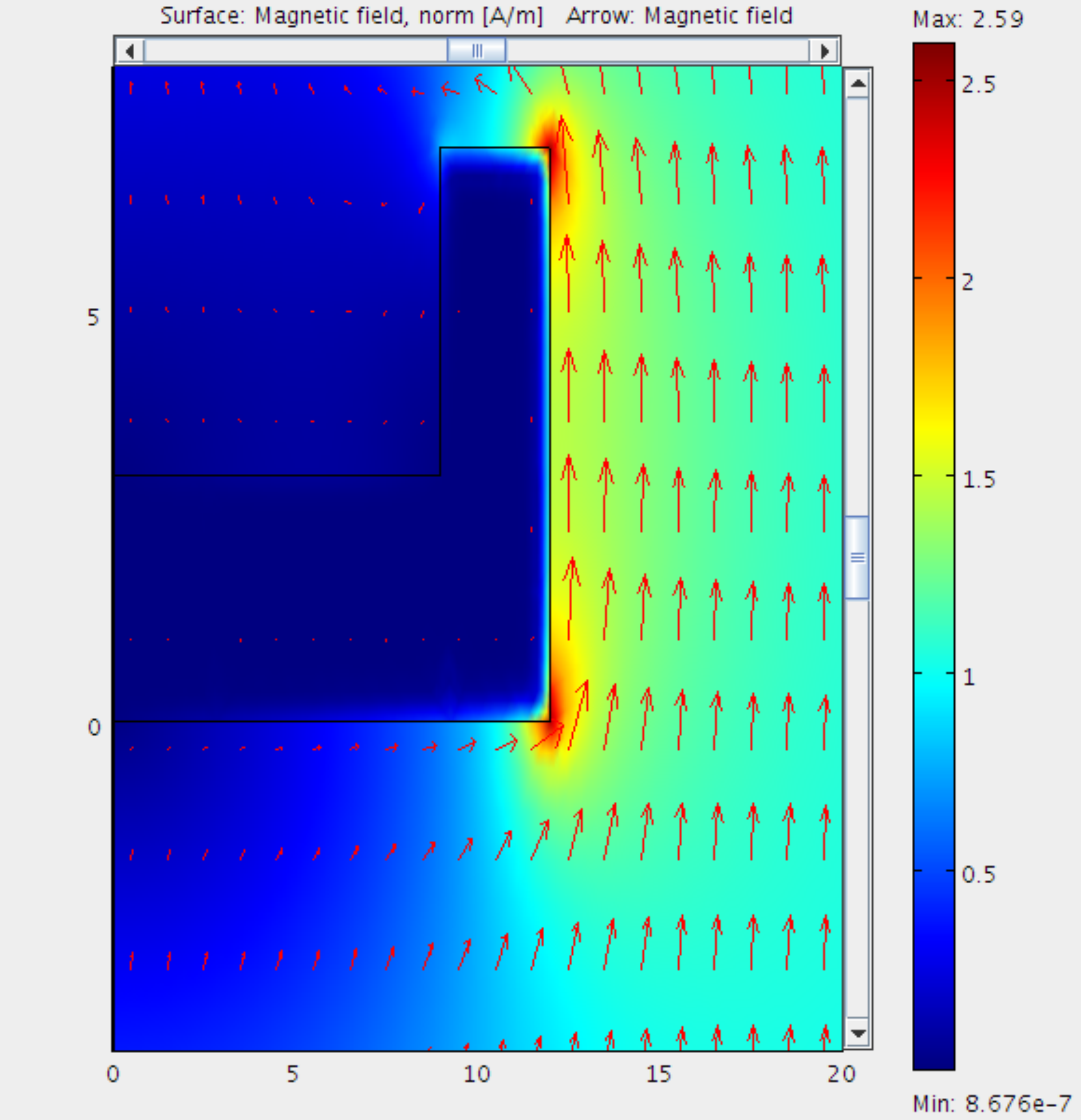} }}
\caption{Magnetic induction field direction (arrows) and modulus (in color code) for a cup shield. }
\label{cup4}
\end{figure}

\begin{figure}
\centerline{
\resizebox{6 cm}{5 cm}{\includegraphics{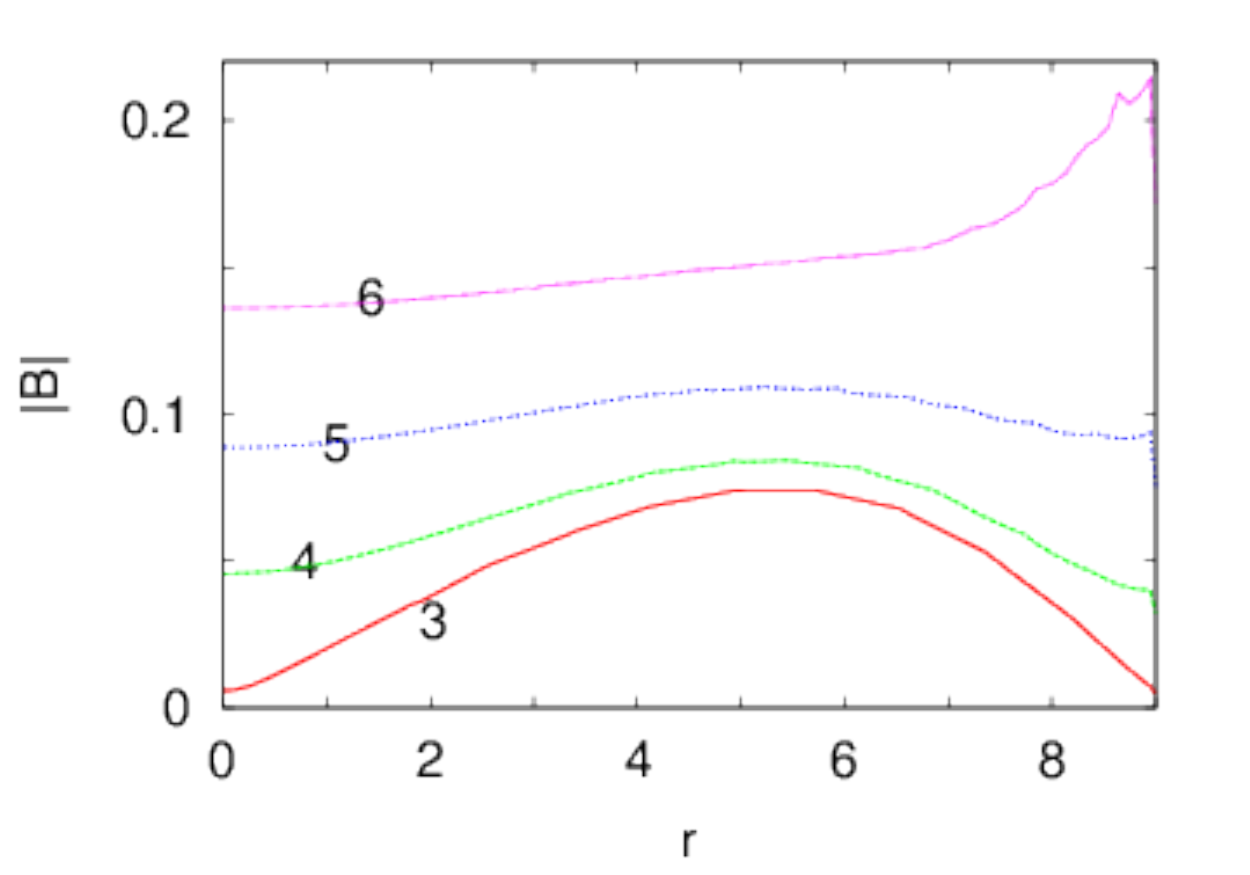}}
\resizebox{6 cm}{5 cm}{\includegraphics{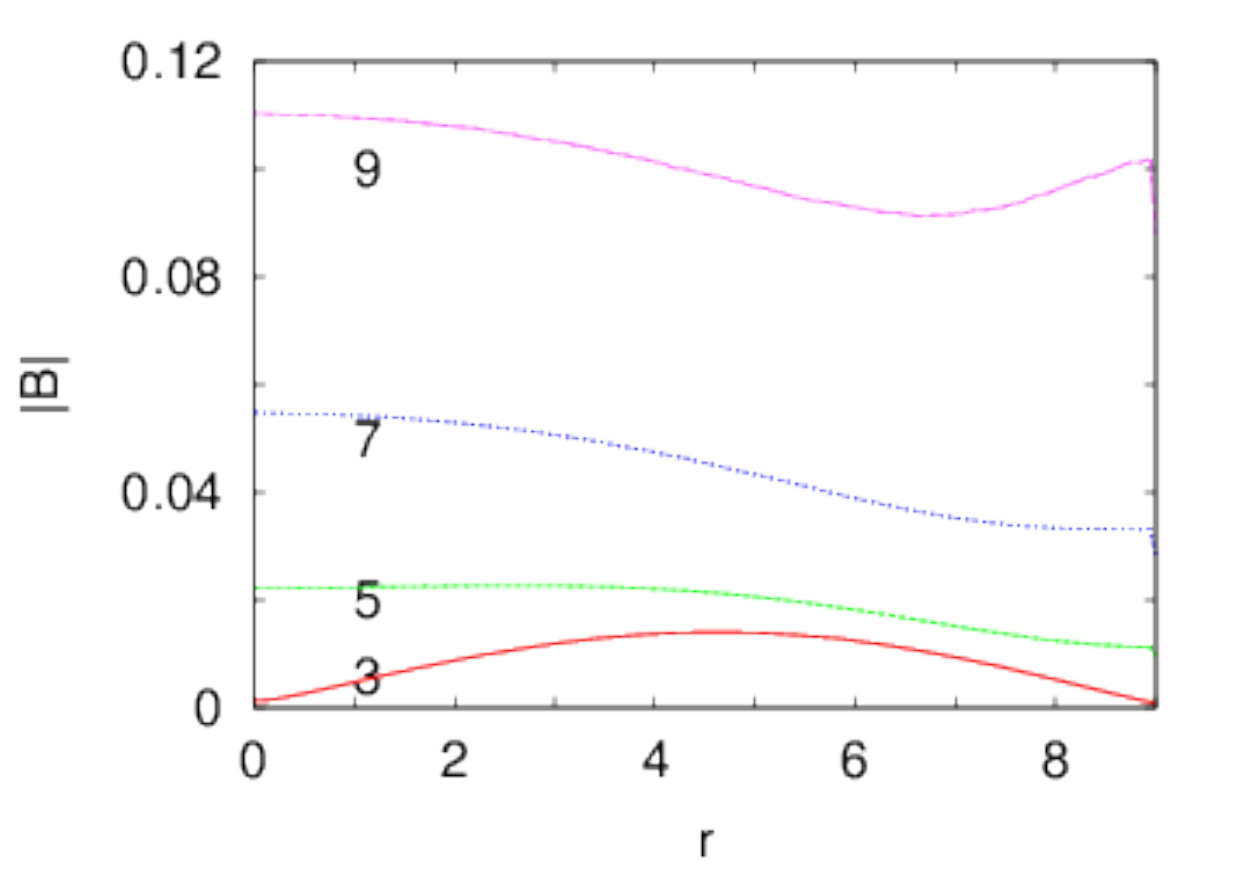}}
}
\caption{Magnetic induction field modulus inside the cup as a function
of $r$ for different values
of $z$. Two different cup depths are shown, $h=4$ mm (left panel)
and $h=8$ mm (right panel). The values of $z$ are indicated on the figures.
}
\label{cup45}
\end{figure}

This magnetic flux distribution allows to design a magnetic screen. This 
design is a problem of shape optimisation which is difficult to solve in general. A simplification
is to assume a shape dependent on a parameter and minimize a criterion with
respect to this parameter. The study done on the disk guides us towards
an ideal geometry. In particular we want to avoid the regions where the field
is reinforced and we want the screening field to remain aligned and opposite
the applied field in the screening region. From our calculations and the
results reported in the literature\cite{denis418} we can rule out 
a cylinder which would
exhibit field reinforcement inside the screening region. Instead, a good candidate
would be a screen shaped as a cup, see Fig. \ref{cup}. Inside 
the cup, the field lines
will cause the screening field to be opposite to the applied field. 
Then one would minimize the magnetic field $\mathbf{B}$ in
a given domain $\Omega$. One could also minimize the magnetic energy
in $\Omega$
\be\label{crit_b}
{\cal E} = {1 \over B_0^2} \int_\Omega \mathbf{B}^2 dr dz =
{1 \over B_0^2} \int_\Omega dr dz
\left[ A_z^2 + ( {1\over r}(rA)_r)^2   \right]  . \ee
To illustrate the procedure we compute the solution for the
cup and change the parameter $h$. The domain $\Omega$ is obviously 
a subset of the
cup interior. The numerical procedure is slightly different than for
the disk since the geometry does not have mirror symmetry. Instead we
apply the same boundary condition (\ref{bc2}) at the two extremities
$z= \pm Z$.
We have chosen $a=3~ mm, r_i=12~ mm~,r_e=r_i+a$ and three different
values of the cup depth $h=2~ mm,4~ mm$ and 8 ~$mm$. The sizes were chosen just to demonstrate the object feasibility: of course in a shield fabrication they can be rescaled in order to meet the experimental constraints.
We take $B_0=1$
so as not to scale the field in equation (\ref{crit_b}). Since the problem is linear, also in this case the unit is arbitrary. Fig. \ref{cup4}
shows the magnetic field for a cup where $h=4$ mm.
The vector field $\mathbf{B}$ is drawn and its modulus is given by the
color code. Notice the strong reinforcement at each edge of the cup.
The dark region (dark blue online) confirms that the field is very small
in the interior of the cup.
As $h$ is increased, the field is reduced inside the cavity. Fig. \ref{cup45}
shows the field in the cavity ($r \le 9$ mm) for a given $z$ as
a function of $r$.
The left panel corresponds to $h=4 ~mm$ and $z=3 ~mm,~4 ~mm,~5 ~mm$ and 6 $mm$
while the right
panel is for $h=8~mm$ and $z=3~mm,~5~mm,~7~mm$ and 9 $mm$. We see that
for $z < a + 2 ~mm$ ($2 ~mm$ above the bottom of the cup) and 
for $h=4 ~mm$ we have
$$ 5 \%  \le {B \over B_0} \le 10 \% . $$
As expected, increasing the height of the cup reduces the field
inside the cup. 
When $h=8 ~mm$ , and $ z < a +4 ~mm$ we have
$$ 1 \%  \le {B \over B_0} \le 5 \% . $$
Therefore increasing the cup depth, one can completely suppress the
magnetic field to a given tolerance. We can then realize a suitable
magnetic field screen.

\vspace{0.4cm}

In summary, we have shown that the Maxwell/London model is
suitable to describe the magnetic field redistribution induced by
a superconducting sample.
This approach was validated by comparing the numerical solutions
to the values of the induction field measured above disk-shaped MgB$_2$ samples.
At T = 20 K, the agreement is very good for external applied field
lower than 0.4 T. This quantitative agreement between experimental results
and a simple model is rare in the literature.

The study indicates that the model can be used also above the
lower critical field, provided that the penetration of the flux lines
inside the sample gives a negligible contribution to the field
values outside the superconductor itself.

The model has no adjustable parameters, since the London penetration
length is a characteristic of the superconducting material
used in the experiment and is introduced a priori. This approach
can be used for superconductors of whatever shape; it also applies
when the external field is inhomogeneous.

Starting from these results, we demonstrated on a cup geometry, how to
design an efficient magnetic field screen by minimizing the magnetic
energy in a given region.
The simplicity
of the direct problem allows to solve this minimization problem easily
and therefore find a cup height $h$ so that the average field inside
a sub-region of the cup interior is below the tolerance.
This is the basis for the design of efficient magnetic field screens.

\section{Acknowledgements}

J.G. C. thanks Michael Sigal for very helpful discussions.
The authors are grateful to the Centre de Ressources Informatiques de Haute
Normandie where most of the calculations were done.
J. G. C. thanks the Department of Mathematics of the University of
Arizona for its hospitality during a sabatical visit.

\end{document}